\documentstyle[twoside,fleqn,espcrc2,epsf]{article}

\begin{document}

\title{Towards the Solution of the Solar Neutrino 
Problem
\thanks{Invited talk presented at the Neutrino-98 conference,
Takayama, Japan, June  1998} 
}
\author{A. Yu. Smirnov\\
{\ }\\
The Abdus Salam International Center 
of Theoretical Physics, 
34100 Trieste, Italy 
\thanks{On leave of absence from INR RAN, Moscow.~~~~~~~~~~~~~~
}
}

\begin{abstract} 
We discuss 
various  aspects of the solar neutrino spectrum distortion and time
variations of  fluxes. 
(i) Oscillations of neutrinos which cross the mantle 
and the core of the
Earth can be parametrically enhanced. The parametric effect gives 
correct physical  
interpretation of the calculated day-night asymmetry. 
(ii) Solution of the $\nu_{\odot}$-problem 
in   schemes with three  and  more neutrinos   
which accommodate explanations of 
other neutrino anomalies, in particular,  the
atmospheric
neutrino anomaly, can lead to   complicated distortion 
of the boron neutrino spectrum. 
(iii)  The study of correlations between time (seasonal or day-night)
variations 
and spectrum distortion  will help to 
identify the solution of the $\nu_{\odot}$-problem.  
 
\end{abstract}

\maketitle

\section{Introduction}

Specific time variations of signals and   
distortion of the energy spectrum  
(along with the  charged to neutral current events ratio) 
are the key signatures of the neutrino 
physics solutions of the solar neutrino problem.  
Preliminary SuperKamiokande (SK)  data \cite{SK} 
indicate that the effects (if exist)
are not strong: $(1 - 2)\sigma$, {\it i.e.}  
at the level of present sensitivity.  
Study of   
correlations between  time variations and  
distortion of the spectrum   strengthens a  
possibility of  identification of  the solution. 
In this connection, I will discuss 
some aspects of the time variations of signals (sect.~II), 
distortion of the energy spectrum (sect.~III) and 
correlation between time variations and spectrum distortion 
(sect.~IV).

\section{What Happens  With  Neutrinos Inside the  Earth?}

The matter of the Earth can modify properties of  
solar, atmospheric and  supernova neutrinos. 
Numerical calculations 
have been performed in a number papers previously 
\cite{earth}, however,   physics of the effects 
has been understood only recently.

The density profile of the Earth has two main structures: 
the core and the mantle. Density changes  slowly 
within the mantle and the core but it jumps sharply 
by a factor of two at their border. It is known for a long while that 
in the first approximation one can consider the mantle and the core as
layers with constant density.  
Neutrinos arriving at the detector at  zenith angle  
$\cos \Theta > -0.84$ cross the mantle only.  
For $\cos \Theta < -0.84$, neutrinos cross three layers: mantle, core
and again mantle. 

Let us introduce $\Phi_m$ and $\Phi_c$  -- 
the oscillation phases  acquired by neutrino
in  the mantle (one layer) and in the core of the Earth:   
\begin{equation}
\Phi_i = 2 \pi \int^{L_i} \frac{dL}{l_i} \approx \int^{L_i}
 d L ~ \Delta H_i~,~~~i = m, c,  
\label{phase}
\end{equation}   
where $l_i  = 2\pi/\Delta H_i$ is the oscillation length in matter, 
and  $\Delta H_i$ is the level splitting (difference of the
eigenvalues of two neutrino states). 
In the layer with constant density: $\Phi_i = \Delta H_i L_i$.

In  \cite{LS} it was realized that for neutrinos which cross both 
the mantle and the core of the Earth 
the equalities 
\begin{equation}
\Phi_{m} \approx \Phi_{c}  \approx \pi
\label{equality}
\end{equation} 
can be approximately satisfied, and this  
leads to  significant enhancement of oscillations.  
(The phases in both layers of mantle are obviously equal.) 
The  transition probability can reach 
\begin{equation}
P^{max}  =  \sin^2 (4 \theta_m - 2\theta_c), 
\label{probab}
\end{equation}  
where $\theta_m$ and $\theta_c$ are the mixing angles in the mantle and
the core respectively. 
$P^{max}$   can be much larger than  $\sin^2 2 \theta_m$ and   
$\sin^2 2 \theta_c$ which correspond to maximal oscillation effect in one
density layer. 

This is a kind of enhancement of oscillations 
which has been introduced by Ermilova et al., \cite{ETC} and  
Akhmedov \cite{Akh} 
(see also \cite{KS}) and  called the 
parametric enhancement of neutrino oscillations.  
The parametric enhancement occurs when the 
parameter of system (the density in our case) changes  
periodically and the period, $r_f$,  coincides with period 
of system.   

The parametric enhancement of  oscillations  is due to certain
synchronization 
of  oscillation effects in the mantle and in the core.
The frequencies  of oscillations are different
in the core and in the mantle. The enhancement occurs
when the  frequency change is synchronized with the frequency itself.

The condition (\ref{equality}) means 
that the size of the layer, $L$,
(in mantle or core) coincides with
half of the oscillation length: $L = l_M /2$. 

In the approximation of constant densities in the mantle 
and the core the resonance condition for phases
(\ref{equality}) can be written as 
\begin{equation}
\Delta H_m L_m  = \pi,~~~   \Delta H_c L_c = \pi~.      
\label{length}
\end{equation}
(In general, the phase should be equal 
$\pi (2k  + 1)$, where $k = 0, 1 , 2 ,...$ fixes 
the order of resonance.)

In 1987 E. Akhmedov \cite{Akh} has considered the case of the 
``castle wall" density profile when the period of perturbation 
consists of two layers with constant but different densities. 
The Earth realizes, in a sense,  the case of ``1.5 period".

The enhancement depends on number of periods (perturbations) and 
on the amplitude of perturbations which can be
characterized by ``swing" angle 
$
\Delta \theta \equiv 2\theta_m - 2\theta_c~.
$
For small perturbations,  
large transition probability can be achieved after many periods. 
In the Earth the perturbation is large $\Delta \theta \sim 
2\theta_c~$, and strong effect is realized even for ``1.5 periods".

Physics of the effect can be well understood from the
graphical representation \cite{KS} based on 
analogy of the neutrino evolution with behaviour of   
spin of the electron in the magnetic field. Indeed,
a neutrino state can be described by vector
\begin{equation}
\vec{\nu} = \left( {\rm Re} \psi_{\mu}^{\dagger} \psi_{s}, ~~
{\rm Im} \psi_{\mu}^{\dagger} \psi_{s},~~
\psi_{\mu}^{\dagger} \psi_{\mu} - 1/2 \right) ~,
\end{equation}
where $\psi_{i}$, ($i =  \mu, s$) are the neutrino wave functions.
(The elements of this vector are nothing but  components of the density
matrix.)
Introducing   vector:
\begin{equation}
\vec{B} \equiv \frac{2 \pi}{l_M} (\cos 2 \theta_M,~~ 0 ,~~ \sin 2
\theta_M)~ 
\label{axisb}
\end{equation}
($\theta_M$ is the mixing angle in medium) which corresponds to the
magnetic field,  one gets from the Schr\"odinger-like 
equation for $\psi_{i}$  the  evolution equation  
\begin{equation}
\frac{d \vec{\nu}}{d t} = \left(\vec{B} \times \vec{\nu} \right)~.
\end{equation}

In medium with constant density ($\theta_M = const$),
the evolution consists of $\vec{\nu}$- precession 
around $\vec{B}$:  $\vec{\nu}$ is moves 
according to increase of the oscillation phase, $\Phi$, 
on the surface of the cone with axis $\vec{B}$.
The direction of the  axis, $\vec{B}$,  
is determined uniquely by $2\theta_M$ (\ref{axisb}).
We will denote by 
$\vec{B}_m$ and $\vec{B}_c$ the axis in
the mantle and in the core respectively. 
In fig.~1 we show a 
projection of the 3-dimensional picture on
$
\left( {\rm Re} \psi_{\mu}^{\dagger} \psi_{s}, ~
\psi_{\mu}^{\dagger} \psi_{\mu} - 1/2 \right)
$
plane \cite{LS}.

The cone angle, $\theta_{cone}$  (the angle between $\vec{\nu}$ and
$\vec{B}$)   depends
both on mixing angle and on  the initial state.
If an  initial state coincides with $\nu_{\mu}$, 
the angle  equals  $\theta_{cone} =  2\theta_M$.
The projection  of $\vec{\nu}$ on the axis $z$,
$\nu_{z}$, gives the probability
to find $\nu_{\mu}$ in a state $\vec{\nu}$:
\begin{equation}
P \equiv \psi_{\mu}^{\dagger} \psi_{\mu} =
\nu_z + \frac{1}{2} =
\cos^2 \frac{\theta_z}{2} ~.
\label{thetaz}
\end{equation}
Here $\nu_{z} \equiv 0.5 \cos \theta_z$,
and  $\theta_z$ is the angle between $\vec{\nu}$ and the axis $z$.

Let us consider an  evolution of the neutrino  which
crosses the mantle, the core and then again the mantle and for which
the  resonance condition  (\ref{equality})  is fulfilled.
In the fig.~1,   
$2\theta_c <  2\theta_m < \pi/2$,   
so that both axes 
$\vec{B}_m$ and $\vec{B}_c$ are in the first  
quadrant. (Actually, such a situation corresponds to 
mixing above the resonance $2\theta_c >  2\theta_m > \pi/2$, 
when  the axes are in the second  quadrant.  In fig.~1 
for convenience of presentation we made redefinition  
$2\theta_c \rightarrow \pi - 2\theta_c$,    
$2\theta_m \rightarrow \pi - 2\theta_m$ which does not change   
result.)  
The initial state, $\vec{\nu}(1)$,  
coincides with flavor state, {\it e.g.},   $\nu_{\mu}$. 
(The picture corresponds to $\nu_{\mu} - \nu_{s}$  mixing   
considered in \cite{LS}.)
\begin{figure}[htb]
\hbox to \hsize{\hfil\epsfxsize=7cm\epsfbox{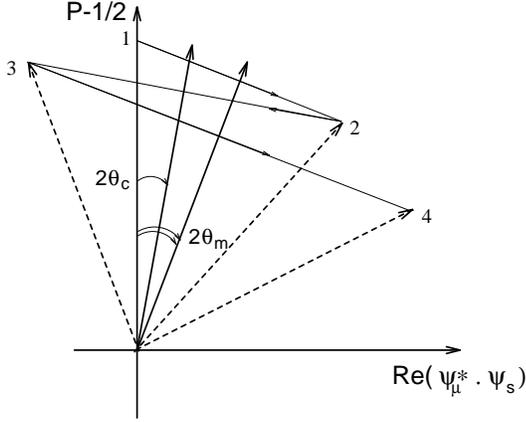}\hfil}
\caption{ Parametric enhancement of the $\nu_{\mu} \leftrightarrow \nu_s$
oscillations inside the Earth.   
Graphical representation of  evolution of the neutrino state; 
the case of parametric resonance. States of neutrino at the borders 
of the layers   are shown  by dashed 
vectors; the cone axes are shown by solid vectors. 
} 
\end{figure}
Neutrino first propagates in the
mantle and this corresponds to $\vec{\nu}$ precession
around  $\vec{B}_m = \vec{B}(2\theta_m)$.  
At the border between the mantle and
the core the neutrino vector is in position $\vec{\nu} (2)$
(which corresponds  to  phase  acquired 
in the mantle,  $\Phi_m = \pi$).  
At the border the mixing angle changes  
suddenly: $\theta_m \rightarrow \theta_c$. 
In the core,   $\vec{\nu}$   precesses around new position of
axis,  $\vec{B}_c \equiv \vec{B}(2\theta_c)$,  with initial condition
$\vec{\nu}(2)$.  At the exit from the 
core,  $\vec{\nu}$   will be in position $\vec{\nu}(3)$. When
neutrino enters  the mantle again, the value of mixing angle
jumps back:
$\theta_c \rightarrow \theta_m$.
In the second layer of mantle,   $\vec{\nu}$
precesses around     $\vec{B}_m$ again.  
At the detector the neutrino vector will be in
position $\vec{\nu}(4)$. After each jump of  density the cone 
angle increases by the value of  
``swing" angle 
$
\Delta \theta \equiv 2\theta_m - 2\theta_c~, 
$
thus enhancing the oscillations. 
According to fig.~1,  a projection of
$\vec{\nu}(4)$ on the axis $z$ equals
$$
\theta_z = 
2\theta_m + 2\theta_m  + 2\Delta \theta  =
2(4 \theta_m - 2 \theta_c )~.
$$
Inserting this into (\ref{thetaz}) we get the survival probability
$\cos^2 (4 \theta_m - 2 \theta_c)$ which reproduces result in 
(\ref{probab}).
\begin{figure}[htb]
\hbox to \hsize{\hfil\epsfxsize=7.5cm\epsfbox{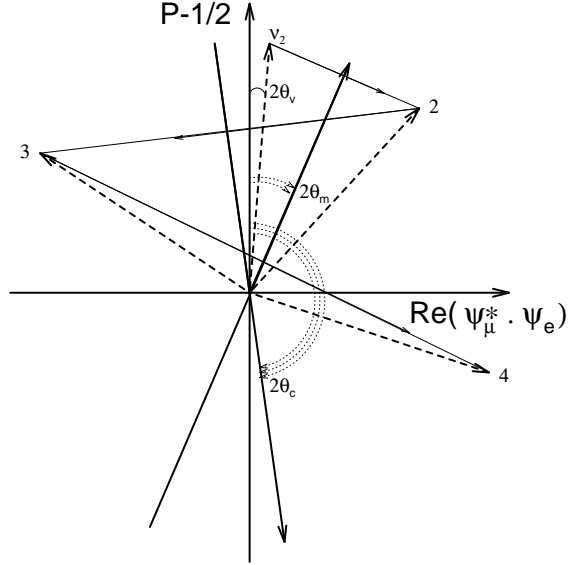}\hfil}
\caption{ Parametric enhancement of the $\nu_2 \rightarrow \nu_e$ 
oscillations inside the Earth.   
Graphical representation of  evolution of the neutrino state;   
the case of parametric resonance.  
}
\end{figure}

In \cite{LS}  the parametric enhancement  
has been applied to $\nu_{\mu} \leftrightarrow \nu_{s}$ oscillations of  
atmospheric neutrinos. 

It was realized by Petcov \cite{P} that the conditions (\ref{equality}) 
are fulfilled for solar neutrinos leading to appearance of the peak 
in the regeneration probability.  
This allows one to get 
correct {\it interpretation} of the dependence of the probability 
on energy found in a number of papers before \cite{earth}. 
It gives correct understanding of the enhancement mechanism.    

Notice that in \cite{P} the condition 
(\ref{equality}) written in the form (\ref{length}) 
has been  {\it renamed}  
by ``oscillation length resonance" and  the enhancement due to 
the condition (\ref{length}) is considered as a new 
effect which differs from that discussed in \cite{ETC,Akh,KS,LS}. 
This change of the name  is unjustified. Indeed,   
the eq. (\ref{length}) is the condition on products of 
inverse oscillation length and width of the layer, that is,  
on the oscillation {\it phases}. 
The resonance associated with equality of phases is the 
parametric resonance.  On the other hand,  
the MSW resonance can  be considered as ``the oscillation 
length resonance": in the MSW resonance   the oscillation
length coincides for small vacuum mixing with refraction length.  

Detailed interpretation of the effect in terms of  the parametric
resonance  has been given  in \cite{Akh1}.

In the case of solar neutrinos the survival probability 
(due to the averaging and lost of coherence) depends on 
the transition probability $\nu_2 \rightarrow \nu_e$ inside the 
Earth, where $\nu_2$ is the heaviest mass eigenstate: 
\begin{equation}
P \approx (1 - 2P_{\odot}) P_{2e}~. 
\label{reg}
\end{equation}
Here $P_{\odot}$ is the $\nu_e$ survival probability 
inside the Sun.

Graphical representation of the evolution of the 
solar neutrinos inside the Earth in the case of parametric resonance  
is shown in fig.~2. 
Now $2\theta_c > \pi/2$  and  $2\theta_m <\pi/2$,  that is,  the axis 
$\vec{B}_m$ is in the first and in the third quadrants,  whereas 
$\vec{B}_c$ is in the second and in the fourth quadrants.
Such a situation corresponds to neutrino energies between the 
MSW resonance energies in the core and in the mantle. 
(It is easy to show that when 
$2\theta_m < 2\theta_c  <\pi/2$ the oscillations are suppressed.) 
The initial state is $\vec{\nu}(1) = \nu_2$.    
Neutrino vector  $\vec{\nu}$ first precesses 
around  $\vec{B}_m$ and at the border between the mantle 
will be in position $\vec{\nu} (2)$. Then 
in the core, $\vec{\nu}$   precesses around    
$\vec{B}_c$,  with initial condition $\vec{\nu}(2)$, 
and at the exit from the core   
$\vec{\nu}$  turns out to be  in position $\vec{\nu}(3)$. 
In the second layer of mantle, the vector  $\vec{\nu}$
precesses around  $\vec{B}_m$  
with initial condition: $ \vec{\nu}=  \vec{\nu}(3)$, 
and at the detector it  will be in
position $\vec{\nu}(4)$. According to fig. 2,  a projection of
$\vec{\nu}(4)$ on the axis $z$ equals
\begin{equation}
\theta_z = 2(4 \theta_m - 2 \theta_c ) - 2\theta~, 
\end{equation} 
and consequently,  
$P_{2e} = \sin^2 (4 \theta_m - 2\theta_c - \theta)$ \cite{P},  
where the difference from (\ref{probab}) is related 
to difference in the initial state. 

One can see from figs.~1 and 2 that enhancement considered in 
\cite{LS} for $\nu_{\mu} - \nu_{s}$ oscillations and the one 
in \cite{P} for $\nu_{2} - \nu_{e}$ are of the same nature: 
the swing of axes  leads to an  enhancement of oscillations. 
The difference is
in the initial state and in inclination of the swing angle.   

Maximal transition probability (\ref{probab}) can be  achieved when the
parametric resonance condition
is fulfilled exactly. The oscillation phases are functions of the
neutrino energy and the zenith angle $\Theta$, and the two
resonance conditions
$\Phi_c (\Theta, E) = \pi $,
$\Phi_m (\Theta, E) = \pi $
can be satisfied only for certain (resonance) values
$\Theta_R$ and $E_R$. Deviations from $\Theta_R$ and $ E_R$
weaken the  enhancement. Thus
the parametric resonance  leads to  appearance of the
peak (parametric peak) in the energy  or/and  zenith angle dependence
of the transition probability.
The width of the parametric peak
is inversely proportional to  number of periods of  
density perturbation:
$ \propto 1/n$  \cite{KS}. (The bigger the number of periods the
sharper the
synchronization condition.) In the case of the Earth the number of periods
is small,  $n  \sim 1.5$, which means that
the width of the peak is of the order one.
Here the enhancement occurs even for  significant detuning.
\begin{figure}[htb]
\hbox to \hsize{\hfil\epsfxsize=8.5cm\epsfbox{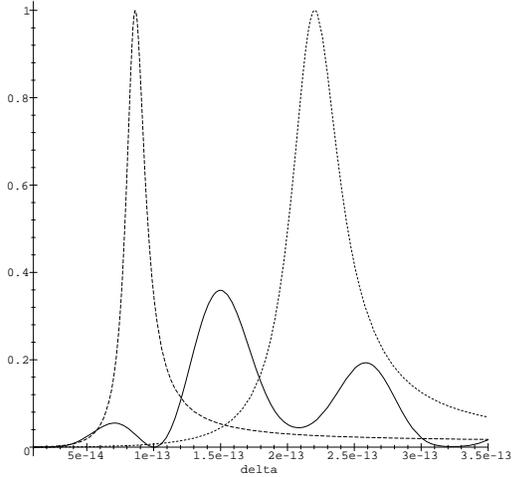}\hfil}
\caption{Transition probability for $\nu_e - \nu_{\mu}$ 
oscillations in the Earth (solid curve) 
as the function of $delta \equiv \Delta m^2/ 4E$. Also shown are 
$\sin^2 2\theta_c$ (dashed curve) and $\sin^2 2\theta_m$ 
(dotted curve); vacuum angle: $\sin^2 2\theta = 0.01$, 
the zenith  angle $\cos \Theta =  - 0.88$.  
(From \cite{Akh1}).  
}
\end{figure}

The probability $P_{2e}$ (as well as $P(\nu_{\mu} \rightarrow 
\nu_e)$)  \cite{earth} has  
rather complicated structure  with three large peaks: 
two of them correspond to the MSW resonance enhancement 
of oscillations in the core and in the mantle. The third peak   
is between the MSW peaks and its height is bigger than 
$\sin^2 2\theta_m$ and  $\sin^2 2\theta_c$ at  the peak energy 
fig.~3.  
The appearance of this third peak associated with resonance condition 
(\ref{equality}) is the consequence of parametric enhancement. 
Notice that  
certain interplay of the oscillation effects in the mantle and in the
core leads not only to appearance of the parametric peak but it also 
modifies  the MSW peaks in the mantle and in the core. 
The MSW  peaks become  suppressed in comparison with 
peaks from only one layer (core or mantle).

Although the parametric enhancement can be rather strong: 
$P_{2e} \sim 1$, the regeneration effect turns out to be  
suppressed  by factor $(1 - 2P_{\odot})$ (\ref{reg}).  
Recent changes in the solar model  predictions 
\cite{BP98,TC} indicate that 
the suppression can be even stronger than it was supposed before. 
Indeed, the predicted flux of the boron neutrinos 
is now smaller (due to smaller cross section of $p Be$ 
reaction). This means that   suppression of the 
boron neutrino flux due to oscillations should be weaker. 
We get $P_{\odot} \sim 0.5$ for the 
neutrino energy $E \sim 10$ MeV -- in the center of the 
detectable region.

\section{Beyond the  Solar Neutrino  Problem}


The solar neutrino problem should be considered 
in general particle physics context which 
allows one also to accommodate solutions of 
other neutrino anomalies, and   
first of all,   the atmospheric neutrino anomaly 
whose oscillation interpretation 
has received strong confirmation \cite{kajita}.  
In fact, the results on atmospheric neutrinos  make 
even more plausible  
the solution of the solar neutrino problem 
in terms of neutrino mass and mixing.

Clearly, the same oscillation channel can not explain 
both the solar neutrino and the atmospheric neutrino anomalies. 
One should consider mixing of three or even more 
neutrino species. 
This can have some impact on  solutions of the 
$\nu_{\odot}$-problem. 
In particular,  one may expect additional  modifications of the neutrino
energy spectrum. 
On the other hand,   the solution 
of the $\nu_{\odot}$-problem may shed some light on the 
origin of other neutrino anomalies.

The  distortion can be  characterized by  a sole
{\it slope parameter}  $s_e$  \cite{rosen}   defined as:
\begin{equation}
\frac{N_{osc}}{N_0} \approx R_0  + s_e T_e  ,
\label{slope}
\end{equation}
where $N_{osc}$ and $N_0$ are the numbers of events with
and without oscillations correspondingly,
$R_0$ is a constant,
$T_e$ is the recoil electron energy in MeV,
$s_e$ is in the units  MeV$^{-1}$.  
\begin{figure}[htb]
\hbox to \hsize{\hfil\epsfxsize=8cm\epsfbox{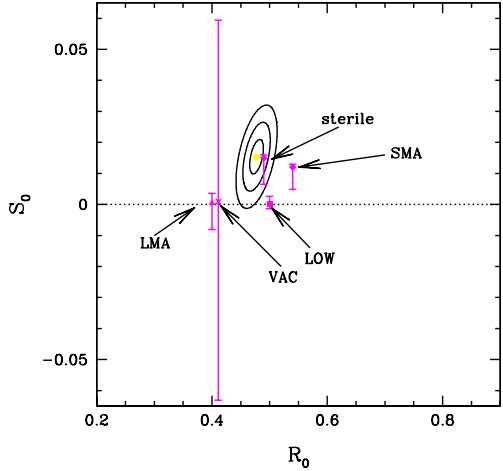}\hfil}
\caption{Deviation from an undistorted energy spectrum. 
The points with error bars show predictions from five 
possible 2$\nu$ - solutions: 
``SMA" stands for small mixing angle MSW conversion $\nu_e \rightarrow
\nu_{\mu}$, 
``sterile" is the small mixing angle MSW conversion $\nu_e \rightarrow
\nu_s$, VAC is the ``just-so oscillations", LMA is the 
large mixing angle MSW solution and LOW is the 
large mixing angle MSW solution with low $\Delta m^2$. The points 
correspond to the best fit points of the total rates in four experiments. 
The ellipses show $1\sigma$, $2\sigma$ and  $3\sigma$ regions allowed 
by SK data. The errors in $R_0$ are large (not shown) so that 
all solution cross the ellipses in the horizontal scale. (From
\cite{bks}.) 
}
\end{figure}
In fig. 4 we show the slope parameter  
predicted by  different two neutrino solutions of the
$\nu_{\odot}$-problem \cite{bks}. 
The dots correspond to the best fit points of the 
total rates. The ellipses show the experimental result. 
Clearly, at the moment it is impossible to 
make discrimination among solutions.  

Let us describe some possibilities beyond simple two neutrino case.

1. In the three neutrino schemes which solve both the solar and 
the atmospheric neutrino problems 
there is the hierarchy: 
$\Delta m^2_{12} \ll \Delta m^2_{13}$. In this case the heaviest state 
``decouples"  from dynamics of the rest of system (leading to the
averaged oscillation result) 
and the survival probability can be written as 
$$
P = \cos^4 \theta_{e3} P_2 + \sin^4 \theta_{e3}~, 
$$
where $\theta_{e3}$ describes the admixture of the 
$\nu_e$ in the heaviest state, and $P_2$ is the two neutrino 
survival probability which is characterized by $\Delta m^2_{12}$ 
and $\sin^2 2\theta_{12}$. For $\Delta m^2_{13} > 2 \times 10^{-3}$ 
eV$^2$ the BUGEY \cite{BUGEY} and CHOOZ \cite{chooz} 
experiments 
give strong bounds on $\theta_{e3}$,  and therefore 
corrections due  presence of the third neutrino 
are small. For $\Delta m^2_{13} <  10^{-3}$ eV$^2$, 
the mixing can be large thus leading to strong modification of the 
probability. Notice,  however, that these changes 
do not improve the fit of the solar neutrino data. 
For small $\theta_{e3}$, the  solutions of these two 
problems essentially decouple \cite{giun}. 

2. All three active neutrinos can be involved in the solar 
neutrino oscillations. This possibility can be naturally realized in the
so called Grand Unification (GU) scenario \cite{babu}.  
Neutrino masses are generated by the 
see-saw mechanism; the  neutrino Dirac mass matrix is 
similar to the mass matrix of the upper quarks at GU scale;  the Majorana
mass matrix of the RH neutrinos has weak mixing and  
linear mass hierarchy with  the heaviest 
eigenvalue at the GU - scale. This scenario 
predicts naturally $\Delta m^2_{13} \sim 10^{-5}$ eV$^2$ -- 
in the range of the MSW solution of the solar neutrino problem 
and $\Delta m^2_{12} \sim 10^{-10}$ eV$^2$ in the ``just-so"  
oscillation region. It also leads to  relatively large 
$\nu_{e} - \nu_{\mu}$ mixing. The solar neutrinos undergo both 
the $\nu_{e} - \nu_{\tau}$ resonance conversion and 
the $\nu_{e} - \nu_{\mu}$ oscillations on the way from the 
Sun to the Earth. The interplay of both effects  results in  a 
peculiar (oscillatory) distortion of the boron neutrino 
energy spectrum \cite{Tak}.    
The corresponding distortion of the recoil energy spectrum 
is shown in fig.~5. 
\begin{figure}[htb]
\hbox to \hsize{\hfil\epsfxsize=8cm\epsfbox{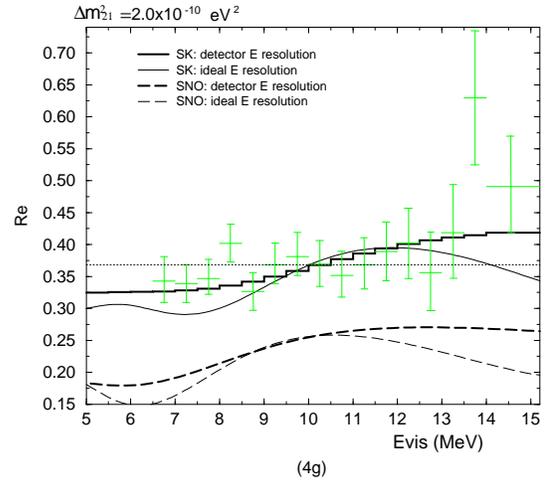}\hfil}
\caption{The expected distortion of the recoil electron
energy spectrum in the SuperKamiokande (solid lines) and SNO (long dashed 
lines) experiments for hybrid solution of the 
$\nu_{\odot}$-problem with parameters: $\sin^2 2\theta_{e \mu} = 0.5$ 
$\sin^2 2\theta_{e \tau} = 6 \times 10^{-4}$, 
$\Delta m^2_{31} = 8 \times 10^{-6}$ eV$^2$ and  
$\Delta m^2_{21} = 2 \times 10^{-10}$ eV$^2$. (From \cite{babu}.)   
}
\end{figure}
Notice that the curve has a kink whose position depends on 
$\Delta m^2$. This may be relevant for  interpretation of the 
SK data. 

3. The atmospheric neutrino problem can be solved by 
oscillations $\nu_{\mu} \leftrightarrow  \nu_s$ which 
involve the sterile 
neutrino. This opens a possibility to rescue small  
{\it flavor}  mixing in lepton sector in analogy with  quark mixing. 
Now  inside the Sun the electron neutrino is converted into 
the mixture of the muon neutrino and sterile neutrino: 
$\nu_2 = \cos \theta_{atm} \nu_{\mu} + \sin \theta_{atm} \nu_s$, 
where $\theta_{atm}$ is the angle responsible for 
deficit of the atmospheric neutrinos. Correspondingly, 
properties of this 
solution of the $\nu_{\odot}$- problem are intermediate between 
properties of solutions based on conversion into 
pure active and pure sterile states. In particular, a 
distortion of the spectrum is stronger than in pure active case but weaker
than in pure sterile case \cite{LS} (fig. 6).  
\begin{figure}[htb]
\hbox to \hsize{\hfil\epsfxsize=8cm\epsfbox{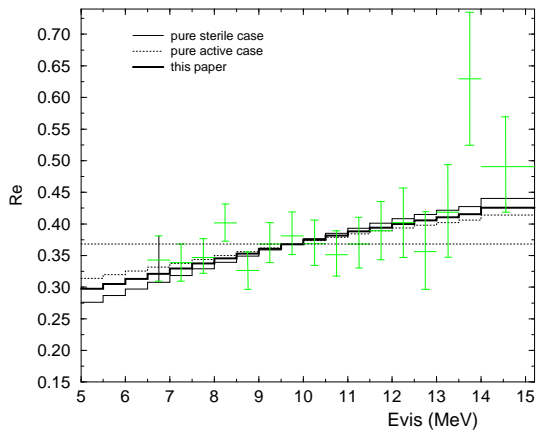}\hfil}
\caption{The expected distortion of the recoil electron energy spectrum in
the SuperKamiokande experiment. The  solid line corresponds to 
pure $\nu_{e} - \nu_s$ conversion, dotted line is for $\nu_{e} -
\nu_{\mu}$,  the bold solid line is for the mixed case 
$\nu_{e} - \nu_{\mu}, \nu_s$ with 
$\Delta m^2 = 5 \times 10^{-6}$ eV$^2$,  $\sin^2 \theta_{12} =8.8 \times
10^{-3}$ and $\sin^2 2\theta_{atm} = 1$. 
}
\end{figure}

4. In the supergravity,  the hidden sector and 
the observable sector communicate via the Planck scale 
($1/M_{P}$) suppressed interactions. 
In particular, a  singlet field $S$ from the hidden sector
may  have the coupling  
$
(m_{3/2}/M_{P}) l H S 
$, where $m_{3/2} \sim 1$ TeV is the gravitino mass, 
$H$ and  $l$  are the Higgs and the lepton doublets
correspondingly. 
This interaction generates the  $\nu-S$ mixing mass term 
\begin{equation}
m_{e s} = \frac{m_{3/2}}{M_{P}} \langle H \rangle  \sim 10^{-4}~~ 
{\rm eV},   
\end{equation}
where $\langle H \rangle$ is the VEV of $H$ \cite{benakli}. 

Consequences of this mixing depend 
on the mass of the scalar, $m_S$. 
It turns out that for  $m_S \sim m_{3/2}^2/ M_{P} \sim 
3 \times 10^{-3}{\rm eV}$ one gets  
$\Delta m^2 \sim 10^{-5}~{\rm eV}^2$ 
and 
mixing angle $\sin^2 2 \theta \sim 10^{-2}$, so that 
the  
$\nu_{e} \rightarrow S$ resonance conversion can solve the 
$\nu_{\odot}$- problem\cite{benakli}. 

If $m_S$ differs from the above value substantially, the other channel, 
{\it e.g.},  $\nu_{e} \rightarrow \nu_{\mu}$,  can give a solution of the
problem. In this case 
the $\nu_{e} - S$ mixing will  modify the two neutrino effect.   
For  $m_S > m_2$ ($m_2 \sim 3 \times 10^{-3}{\rm eV}$), 
the $\nu_{e} - S$  mixing can  lead to 
a  dip in the non-adiabatic edge of the suppression pit at 
$E \sim (m_S/m_2)^2 E_a$, where  $E_a \sim (0.5  - 0.7)$ MeV is the 
energy of the adiabatic edge. 
This will manifest as a  dip in the recoil electron spectrum  
and  can be relevant for 
explanation of the  spectrum observed by the
SuperKamiokande. Also flavor composition of the 
neutrino flux will depend on energy.  
The flux of the beryllium $\nu_e$ neutrinos is  converted mainly 
to $\nu_{\mu}$, whereas boron neutrinos are transferred both to 
$\nu_{\mu}$ and $S$. Correspondingly, an effect of the neutral
currents  is 
larger for low energies. Comparison of signals in BOREXINO and SNO 
experiments will check this effect.


\section{Time  Variations  Versus Distortion}


Existing solutions of the 
$\nu_{\odot}$-problem lead to specific   
correlations between  time variations of signals and 
spectrum distortion. Therefore, using the data on 
spectrum distortion one can
make predictions for time variations and {\it vice versa}. 
A study of 
these correlations  strengthens 
the possibility to identify 
the  solution. 

1. For vacuum oscillation solution there 
is a strict correlation between a
spectrum distortion and the amplitude of seasonal variations of 
neutrino flux \cite{ms-sk}. The seasonal variations are  due to
ellipticity of 
the Earth orbit. The correlation originates from  dependence of the 
oscillation probability $P$ on the neutrino energy and distance to the
Sun. Indeed, the phase of
oscillations is proportional to 
$\Phi \propto L/E$ which gives immediately 
\begin{equation}
\frac{d P}{d L} = - \frac{d P}{d E} \cdot \frac{E}{L}~.  
\label{relation}
\end{equation}
Here $P^{-1}{d P}/{d E}$ is the slope of the neutrino spectrum
distortion. 
According to (\ref{relation}), a positive slope, ${d P}/{d E} > 0$, is
accompanied by decrease 
of probability with distance, so that the seasonal variations due to 
geometrical factor, $L^{-2}$, will be enhanced.  
In the case of negative slope, oscillations will suppress 
the seasonal variations due to geometrical factor.  

The correlations can be expressed as correlations  between the slope 
parameter for the energy spectrum of the recoil electrons 
(\ref{slope}) and the 
summer-winter asymmetry defined as 
\begin{equation}
A_e \equiv 2 \frac{N_{W} - N_{S}}{N_{SP} + N_{A}} .
\label{assym}
\end{equation}
Here  $N_{W}$,  $N_{S}$ , $N_{SP}$, $N_{A}$
are the  numbers of events detected
from November 20 to February 19, from May 22 to August 20,
February 20 to May 21, from August 21 to November 19
respectively. It is convenient to describe the asymmetry 
due to oscillations  by the parameter 
\begin{equation}
r_e \equiv \frac{A_e}{A_e^0} -1~,    
\end{equation}
where 
$A_0$ is the asymmetry related  to the geometrical factor. 
Obviously,  $r_e = 0$ in the  no-oscillation case; 
$r_e > 0$ ($r_e < 0$) corresponds to enhancement (damping) of  
the geometrical effect. 
Fig. 7 shows the $s_{e}-r_e$ correlation. 
For the best fit value of the  slope (fig. 4)  
we get  $r_e \sim 0.4$, so that  one expects an enhancement of asymmetry. 
This can be checked after 4 - 5 years of the SK operation. 
\begin{figure}[htb]
\hbox to \hsize{\hfil\epsfxsize=7cm\epsfbox{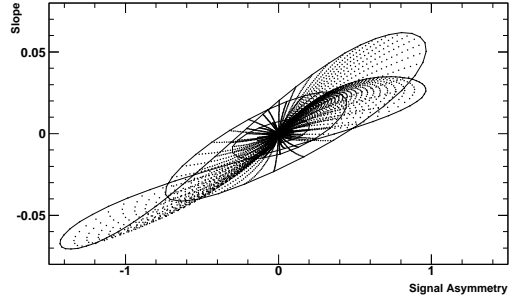}\hfil}
\caption{The slope - asymmetry plot. The points correspond to different
values of $\Delta m^2$ between $10^{-11}$ and $10^{-9}$ eV$^2$, 
and $\sin^2 2\theta$ between 0.25 and 1.00. The solid line shows changes
of the slope and asymmetry with $\Delta m^2$ for maximal mixing.   
}
\end{figure}

2. In the case of the MSW solution there is a correlation between the
day-night asymmetry and spectrum distortion. This helps 
do disentangle the large and small mixing solutions 
of the problem \cite{lisi}. 
For large mixing solution one expects strong day-night
asymmetry and weak distortion of the spectrum. In contrast, for small
mixing solution stronger spectrum distortion is accompanied by weak 
day-night effect. In fig.~8 the distortion of spectrum is characterized by
deviation of the average electron kinetic energy $T_e$ from 
its standard value without oscillations. As follows from the figure the
data  favor a small mixing solution. 
\begin{figure}[htb]
\hbox to \hsize{\hfil\epsfxsize=7cm\epsfbox{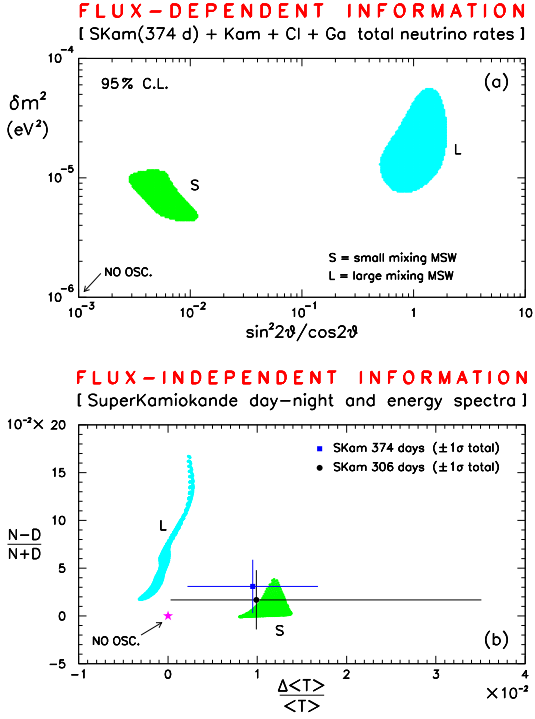}\hfil}
\caption{The day-night asymmetry - spectrum distortion plot. 
The distortion is characterized by the mean kinetic energy 
deviation. In panel (b) the regions show 
the map of the small (S) and large (L) 
mixing solutions at 95 \% C. L. in the mass-mixing plane (panel(a)).  
(From \cite{lisi}.)
}
\end{figure}

3. The correlation of the day-night effect 
and spectrum distortion  allows one 
also to disentangle solutions based on  conversion to active and to
sterile neutrinos. Main difference comes from presence of the 
$\nu_{\mu}(\nu_{\tau})$ contribution to $\nu e$-scattering in the case of
active neutrino conversion. This contribution, being proportional 
to $(1 - P(E))$,  leads to smearing of the spectrum distortion. Therefore 
for the same  values of parameters the distortion is stronger in the
sterile case. In contrast, the regeneration effect is weaker in the
sterile neutrino case. This is related to the fact, that in the 
$\nu_e - \nu_s$ case the effective potential (which describes  matter
effect) is  approximately two times smaller than  
in the $\nu_e - \nu_{\mu}$-case. Thus for $\nu_e - \nu_{\mu}$ conversion  
one expects larger day-night asymmetry and smaller slope, 
whereas  $\nu_e - \nu_s$ conversion leads to larger slope but weaker  
asymmetry. In fig.~9 we show projection of the 
($\Delta m^2, \sin^2 2 \theta$)  regions of
 small mixing  solutions 
onto D/N-asymmetry - slope plot 
which illustrates the correlation \cite{LS2}. The correlation is 
solar model dependent. 
For the model  BP95 \cite{BP95} the regions 
corresponding to two channels of conversion are well separated. 
However in the models with smaller boron neutrino flux 
(see e.g.  \cite{BP98}, \cite{TC}) 
both the slope and the D/N asymmetry become smaller and 
the two regions overlap. The  
identification of solutions (using this correlation) will be difficult. 
 Notice that for small original boron neutrino flux the D/N asymmetry 
is negative 
in  whole region of the $\nu_e - \nu_s$ solution and in  part 
of the $\nu_e - \nu_{\mu}$ region.   
This is related to the fact, that for a small flux a required 
oscillation suppression should be weak, so that the survival probability 
$P_{\odot} > 1/2$ (see (\ref{reg})). Moreover, due to 
additional contribution from $\nu_{\mu}$ the $\nu_e - \nu_{\mu}$ solution 
requires stronger suppression.   
The two solutions can be also distinguished by measurements 
of the neutral current effect in SNO. 
\begin{figure}[htb]
\hbox to \hsize{\hfil\epsfxsize=8cm\epsfbox{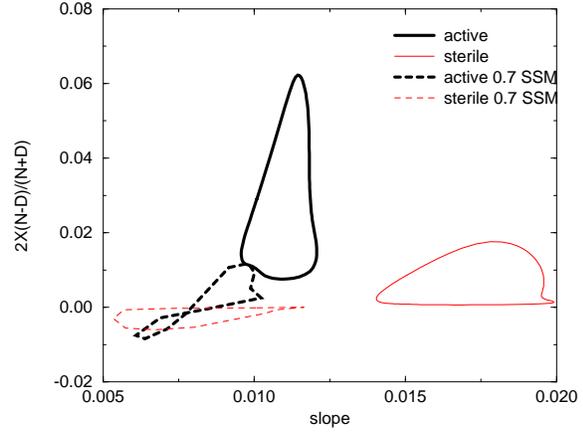}\hfil}
\caption{The slope - D/N-asymmetry plot \cite{LS2}. The regions of
predictions 
of small mixing MSW solutions: $\nu_e - \nu_{\mu}$ (bold lines), 
$\nu_e - \nu_s$ (thin lines). Solid lines correspond to 
solar model BP95, dashed lines are for BP95 model with 
diminished (by factor 0.7) boron neutrino flux. 
}
\end{figure}

\section{Conclusion}


Oscillations of neutrinos crossing the core of  Earth can be
parametrically enhanced. This leads to appearance of the parametric peak
in the oscillation probability as function of neutrino energy. 
The parametric enhancement can be relevant for solar and  atmospheric
neutrinos as well as for neutrinos from supernova. 
Strong enhancement of the regeneration 
probability for solar neutrinos which cross  the core is due to the 
parametric resonance.

Solution of the solar neutrino problem should be considered in wider
particle physics context which allows one to explain,  
{\it e.g.},  the atmospheric 
neutrino problem. Under certain conditions the two problems  
``decouple" and the solution is still reduced to simple two neutrino case. 
However, in a number of schemes one gets modification of the simple two
neutrino effect. This can  manifest as  
complicated distortion of the neutrino (and the recoil electron) 
energy spectrum and also can lead to a peculiar change of the flavor
composition of the solar neutrino flux with energy. 

Precise measurements of spectrum can reveal physics ``beyond 
the solar neutrino problem". One possibility is 
the Planck mass suppressed  couplings of neutrinos with particles from
the hidden sector. 

Different solutions of the solar neutrino problem lead to 
specific  correlations
between the spectrum distortion and time variations
of fluxes. This can be used to distinguish  solutions. 

Recent experimental data and new calculations of the fluxes 
require smaller oscillation effects (smaller 
mixing angles etc.), so that the identification of the solution 
becomes more  difficult.


\begin{thebibliography}{99}

\bibitem{SK} Y. Suzuki, (these proceedings). 

\bibitem{earth} 
S. P Mikheyev, A. Yu. Smirnov, In {\it '86 Massive Neutrinos in 
Astrophysics and in Particle Physics}, Proc. of the 6th Moriond workshop, 
edited by O. Fackler and J. Tran Than Van, p 355. 
J. Bouchez {\it et al}, Z. Phys. C {\bf 32} (1986) 499; 
E. D. Carlson, Phys. Rev D {\bf 34}, 1454 (1986);
M. Cribier, W. Hampel, J. Rich, and D. Vignaud, Phys. Lett. B {\bf 182},
89 (1986);
A. J. Baltz and J. Weneser, Phys. Rev. D {\bf 35}, 528 (1987);
A. Dar, A. Mann, Y. Melina, and D. Zajfman, Phys. Rev. D {\bf 35}, 3607
(1988); 
G. Auriemma, M. Felcini, P. Lipari and J. L. Stone,
Phys. Rev. D {\bf 37}, 665 (1988); 
A. Nicolaidis, Phys. Lett. B {\bf 200}, 553 (1988);
P. I. Krastev and S. P. Petcov, Phys. Lett. {\bf B205}, 84 (1988);
J. M. LoSecco, Phys. Rev. D {\bf 47}, 2032 (1993);
J. M. Gelb, W.-K. Kwong, and S. P. Rosen, Phys. Rev. Lett.
{\bf 78}, 2296 (1997); 
J.N. Bahcall, P.I. Krastev, Phys. Rev. C {\bf 56} (1997) 2839; 
Q. Y. Liu, M. Maris and S. T. Petcov, Phys. Rev. D
{\bf 56}, 5991 (1997); 
M. Maris and S. T. Petcov, Phys. Rev. D {\bf 56},
7444 (1997); 
M. Maris and S. T. Petcov, hep-ph/9803244.

\bibitem{LS} Q. Y. Liu and A. Yu. Smirnov, Nucl. Phys. B524
(1998) 505, hep-ph/9712493; Q. Y. Liu, S. P. Mikheyev and A. Yu. Smirnov,
hep-ph/9803415.

\bibitem{ETC} V. K. Ermilova, V. A. Tsarev and V. A. Chechin, Kr. Soob,
Fiz. [Short Notices of the Lebedev Institute] {\bf 5}, 26 (1986).

\bibitem{Akh} E. Kh. Akhmedov, preprint IAE-4470/1, (1987); Yad. Fiz. {\bf
47}, 475 (1988) [Sov. J. Nucl. Phys. {\bf 47}, 301 (1988)].

\bibitem{KS} P. I. Krastev and A. Yu. Smirnov, Phys. Lett. B {\bf 226},
341 (1989).



\bibitem{P} S. T. Petcov, preprint SISSA 31/98/EP, hep-ph/9805262.

\bibitem{Akh1} E. Kh. Akhmedov, hep-ph/9805272.



\bibitem{BP98} J. N. Bahcall, S. Basu,  M.H. Pinsonneault, 
Phys. Lett. B {\bf 433} (1998), 1. 

\bibitem{TC} A. S. Brun, S. Turck-Chieze and P. Morel, astro-ph/9806272. 


\bibitem{kajita} T. Kajita, these proceedings.


\bibitem{rosen} see {\it e.g.} W. Kwong, S. P. Rosen, Phys. Rev. D 
{\bf 51} (1995) 6159.  

\bibitem{bks} J. Bahcall, P. Krastev and A. Smirnov, hep-ph/9807216.

\bibitem{BUGEY} B Achkar et al., Nucl. Phys. B {\bf 434}, (1995) 503.  


\bibitem{chooz} CHOOZ collaboration, M. Apollonio et al. hep-ex/9711002.

\bibitem{giun} see {\it e.g.} C. Giunti, hep-ph/9802201. 

\bibitem{babu} K.S. Babu, Q.Y. Liu, A.Yu. Smirnov, 
Phys. Rev. D {\bf 57} (1998) 5825, hep-ph/9707457. 

\bibitem{Tak} A. Yu. Smirnov, Proc. of the  Int. symposium 
{\it Frontiers of Neutrino Astrophysics}, Takayama, October 1992 
 Ed. Y. Suzuki and K. Nakamura, (1992) p. 105; 
Q.Y. Liu, S.T. Petcov,  Phys. Rev. D {\bf 56} (1997) 7392; 
B.C. Allanach, G.K. Leontaris, S.T. Petcov,  
Phys. Lett. B {\bf 431} (1998) 98.  

\bibitem{benakli} K. Benakli,  A. Yu. Smirnov,  
Phys. Rev. Lett. {\bf 79} (1997) 4314.  

\bibitem{ms-sk} S.P. Mikheyev, and A.Yu. Smirnov, 
Phys. Lett. B {\bf  429} (1998) 343. 

\bibitem{lisi} G.L. Fogli, E. Lisi, D. Montanino,  hep-ph/9803309.  

\bibitem{LS2}Q. Y. Liu  and A. Yu. Smirnov (in preparation).

\bibitem{BP95} J. N. Bahcall and M. H. Pinsonneault, 
Rev. Mod. Phys. {\bf 67} (1995) 781. 



\end{thebibliography}
\end{document}